\documentclass[conference]{IEEEtran}
\IEEEoverridecommandlockouts
\usepackage{cite}
\usepackage{amsmath,amssymb,amsfonts}
\usepackage{algorithmic}
\usepackage{graphicx}
\usepackage{booktabs}
\usepackage{textcomp}
\usepackage{xcolor}
\usepackage{enumitem}
\usepackage{cite}
\usepackage{comment}

\def\BibTeX{{\rm B\kern-.05em{\sc i\kern-.025em b}\kern-.08em
    T\kern-.1667em\lower.7ex\hbox{E}\kern-.125emX}}

\begin{document}

\title{
{\vspace{-12mm}\textit{\normalsize{This work has been submitted to the IEEE for possible publication. \\ \vspace{-6mm}
Copyright may be transferred without notice, after which this version may no longer be accessible.}}} \\ \vspace{5mm}

Design and Development of a Robust Tolerance Optimisation Framework for Automated Optical Inspection in Semiconductor Manufacturing
\thanks{This knowledge transfer partnership between Ulster University and Elite Electronic Systems Ltd is funded by Innovate UK Knowledge Transfer Network (KTN) and Invest Northern Ireland (Invest NI) [Project Number: 10078007].}
}

% \author{\IEEEauthorblockN{Shruthi Kogileru}
% \IEEEauthorblockA{\textit{School of Engineering} \\
% \textit{Ulster University}\\
% Belfast, UK \\
% s.kogileru@ulster.ac.uk}
% \and
% \and
% \IEEEauthorblockN{Yaxin Bi}
% \IEEEauthorblockA{\textit{Artificial Intelligence Research Centre} \\
% \textit{Ulster University}\\
% Belfast, UK \\
% y.bi@ulster.ac.uk}
% \and
% \IEEEauthorblockN{Kok Yew Ng}
% \IEEEauthorblockA{\textit{Engineering Research Institute} \\
% \textit{Ulster University}\\
% Belfast, UK \\
% mark.ng@ulster.ac.uk}
% }

\author{\IEEEauthorblockN{Shruthi Kogileru\IEEEauthorrefmark{1}\IEEEauthorrefmark{2}, Mark McBride\IEEEauthorrefmark{2}, Yaxin Bi\IEEEauthorrefmark{3}, and Kok Yew Ng\IEEEauthorrefmark{1}\IEEEauthorrefmark{4}}
\IEEEauthorblockA{\IEEEauthorrefmark{1}School of Engineering, Ulster University, Belfast, UK.
Emails: \{s.kogileru, mark.ng\}@ulster.ac.uk}
\IEEEauthorblockA{\IEEEauthorrefmark{2}Elite Electronic Systems Ltd, Enniskillen, UK.
Email: mmcbride@elitees.com}
\IEEEauthorblockA{\IEEEauthorrefmark{3}School of Computing, Ulster University, Belfast, UK.
Email: y.bi@ulster.ac.uk}
\IEEEauthorblockA{\IEEEauthorrefmark{4}School of Engineering, Monash University, Bandar Sunway, Malaysia}}

\maketitle

\begin{abstract}
    Automated Optical Inspection (AOI) is widely used across various industries, including surface mount technology in semiconductor manufacturing. One of the key challenges in AOI is optimising inspection tolerances. Traditionally, this process relies heavily on the expertise and intuition of engineers, making it subjective and prone to inconsistency. To address this, we are developing an intelligent, data-driven approach to optimise inspection tolerances in a more objective and consistent manner. Most existing research in this area focuses primarily on minimising false calls, often at the risk of allowing actual defects to go undetected. This oversight can compromise product quality, especially in critical sectors such as medical, defence, and automotive industries. Our approach introduces the use of percentile rank, amongst other logical strategies, to ensure that genuine defects are not overlooked. With continued refinement, our method aims to reach a point where every flagged item is a true defect, thereby eliminating the need for manual inspection. Our proof of concept achieved an 18\% reduction in false calls at the 80th percentile rank, while maintaining a 100\% recall rate. This makes the system both efficient and reliable, offering significant time and cost savings.
\end{abstract}


\section{Introduction}
The semiconductor industry is one of the fastest growing industries in the world. The assembly of Printed Circuit Boards (PCBs) is a complex process involving the placement of a large number of electronic components of various shapes and sizes on the boards. Surface Mount Technology is a PCB manufacturing technique that has become the popular choice for manufacturers because of its high precision, speed and accuracy \cite{yimin2023industrial}. In the SMT process, first a bare board is smeared with solder paste. Next, pick-and-place machines place the components in their designated locations. The board then passes through high temperatures in a reflow oven where the solder paste melts and connects the components onto the board. Finally, this populated board undergoes quality inspection through an Automated Optical Inspection machine at the end of the production line \cite{Kim}.  

\begin{figure}[t]
    \centering
    \includegraphics[width=\columnwidth]{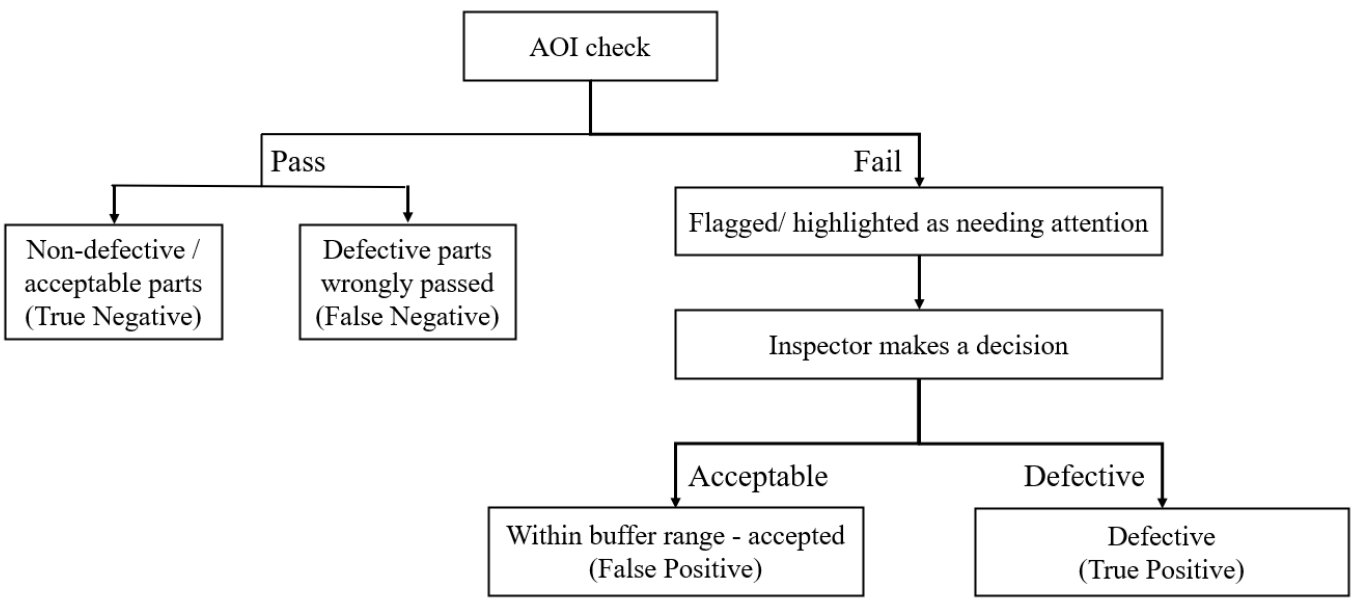}
    \caption{AOI inspection system process.}
    \label{fig:fig1}
\end{figure}

The AOI plays an important role in ensuring high levels of product quality and production efficiency by performing 2D and 3D optical checks to ensure the products comply with the necessary standards. As such, AOI systems have predefined tolerance values set for each part for each type of inspection, i.e. solder, alignment, optical character recognition etc., which the boards are compared against to determine if an inspection is acceptable or defective. See Fig. \ref{fig:fig1}. Any inspection that falls outside of the allowable tolerance would be flagged by the AOI system, after which an inspector would then check the flagged part and make a decision as to whether it is an actual defect or if it is acceptable, with the latter being categorised as a \textit{false flag}. All good inspections are categorised as true negatives, whereas the true defects are considered true positives. False negatives are the true defects that are passed undetected by the AOI, and false positives are the inspections flagged by the AOI for review, but are acceptable as good.

It is of utmost importance to ensure high levels of production efficiency and quality of the products manufactured. To achieve this, there has been a widespread adaptation of Industry 4.0 technologies in the manufacturing industry. Factories are now shifting towards smart manufacturing processes to enable modelling, simulation, and intelligent decision making to improve production capabilities and ensure continuous improvement. The Digital Twin technology is one such leading Industry 4.0 technology that acts as a digital duplication of actual entities by providing simulation capabilities to reflect the physical status of a system in a virtual space \cite{Lim,Manjurul}. 

One of the greatest challenges in the manufacturing industry is striking the right balance between minimising false flags while preserving optimal product quality by preventing genuine defects from passing inspection. This becomes especially critical when working with high-stakes PCBs intended for devices in the medical, defence, and automotive sectors, where unwavering caution is paramount to ensure no real defects go undetected. Therefore, setting the appropriate tolerances in the AOI machine is a crucial step to ensure the quality and reliability of the PCBs produced.  
The most widespread practice in the manufacturing industry is for engineers to determine tolerances based on their knowledge, experience, and intuition. This trial-and-error approach often results in tolerances that are either too lenient or too strict, leading to needless expenditure of time and resources. 

This paper proposes a statistical, data-driven method that enables dynamic recalibration through real-time feedback of tolerance values after each inspection. The system is designed to balance false flag minimisation, process stability, and product and process quality. Optimised tolerance values enable engineers to make informed decisions on adjusting AOI machine settings without relying solely on trial and error.
The proposed method aims to improve production efficiency by reducing false flags and thus freeing up resources, mitigating delays, and enhancing throughput. Fewer false flags also reduce the burden on human inspectors, lowering the risk of genuine defects being overlooked. Moreover, this approach alleviates the reliance on individual engineers’ experience, allowing for a more reliable and data-driven optimisation of tolerance values. 

Section \ref{sec:existingwork} reviews existing work relevant to our proposed system. Section \ref{sec:method} outlines the methodology used to implement the model. Section \ref{sec:results} presents the initial research findings and proof of concept, along with some discussions. Finally, Section \ref{sec:conclusion} concludes the paper.

\section{Backgrounds and Existing Work}\label{sec:existingwork}
While significant research addresses design tolerance optimisation in manufacturing, there is very little focus on production inspection machines such as AOIs for manufactured products. For AOIs inspecting PCBs, each component may demand a distinct tolerance for each type of inspection, leading to thousands of possible scenarios. Consequently, a general paradigm is needed that accommodates inspection of all components, including new introductions that the system has not previously encountered.

In \cite{Res}, We can see that balancing the trade-off between reducing false calls and maintaining quality is very challenging. After comparing against multiple models like CART, Random Forest and ANN, the k-nearest neighbours (kNN) approach was proposed for classifying AOI results, reducing false calls by 32.4\%. This model initially classifies defects and forwards only high-probability cases to the inspector for review. Although this approach places responsibility on the model for genuine defects that might be incorrectly labelled as non-defective, it does not strictly prevent such misclassification, which is risky in safety-critical applications. Furthermore, while the proposed model is theoretically sound, there is no clear methodology for practical implementation.

In \cite{9935366}, the proposed methodology involves transfer learning from pre-trained deep learning models to classify a PCB as defective or not, which is advantageous when the dataset is not extensive enough to train a deep-learning model from scratch. An Xception model with pre-loaded weights from ImageNet was used for feature extraction. The original classification layer was removed, and a new classification layer with a sigmoid activation function was added for binary classification (true or false call). Fine-tuning this model by training the few top layers along with the new classification layer gave a classification accuracy of 91.3\% on evaluation. Along with the accuracy being relatively low, the model does not calculate recall. There is also no implementation methodology provided.

\section{Methodology}\label{sec:method}
After each board inspection, the AOI system acquires and stores data from about 50 parameters, such as length, rotation, offset, and component type. Depending on the nature of the inspection (e.g., solder, tombstone, lead height, or 3D inspection), only the relevant columns needed for our optimisation framework are extracted. These include the part number, the specific PCB models in which the component is used, the inspection parameters, and their corresponding tolerances.

We developed our optimisation system using Python scripts. First, inspection data from the AOI machines was extracted and preprocessed by eliminating rows with missing data. Then, the data was transformed by selecting only the required columns, converting the inspection values into floating point numbers, and storing them in a uniform format.

In a subsequent phase, we evaluate the performance of our model by examining its ability to detect all true defects. Specifically, we identify the five distinct parts with the highest number of confirmed defects. We then partition those defects into a 70--30 split; 70\% for tolerance recalibration, and the remaining 30\% comprising previously unseen data for validation. Restricting the selection to the top five parts ensures sufficient data for both recalibration and validation. We then employ percentiles to compute the updated tolerance thresholds, and the steps to obtain the optimized tolerance are outlined as follows: \vspace{1mm}

\noindent \underline{\textbf{Step 1:}} Retrieve all relevant features and their associated pass criteria for the specified part number. Then, isolate and collect only those measurements categorised as false calls such that
    \begin{equation}
        X = x_1, x_2, x_3, \dots , x_n, \label{eq:eq1}
    \end{equation}
    where $n$ is the total number of measurements. \vspace{1mm}

\noindent \underline{\textbf{Step 2:}} The values in (\ref{eq:eq1}) are then sorted using
\begin{equation}
    X = \text{sort}(X) = \{ x_i \in X \mid x_i \leq x_{i+1},\ \forall i \in [1, n-1] \}.
\end{equation} 

\noindent \underline{\textbf{Step 3:}} The new tolerance is derived using a percentile-based approach. Because the chosen percentile value is adjustable, we can assess the impact of various options via simulations prior to implementing any tolerance changes within the AOI. Hence, the rank of the chosen percentile can be found,
\begin{equation}
    \mbox{Rank} = \frac{p(n-1)}{100} + 1,
\end{equation}
where $p$ is the percentile. \vspace{1mm}

\begin{comment}
\noindent \underline{\textbf{Step 4:}} The linear value corresponding to the rank determined in Step 3 is calculated using $i = \lfloor \mbox{Rank} \rfloor$ and subsequently, $d = \mbox{Rank} - i$,
% \begin{equation}
%     i = \lfloor \mbox{Rank} \rfloor, ~~~~~d = \mbox{Rank} - i,
% \end{equation}
where $i$ is the integer part and $d$ is the decimal part. As a result, the new percentage can be computed using $p = x_i + d (x_{i+1} - x_i)$.
% \begin{equation}
%     p = x_i + d (x_{i+1} - x_i).    
% \end{equation}
By applying these procedures, a newly optimised tolerance value can be obtained \cite{Bornmann}. \vspace{1mm} 
\end{comment}

\noindent \underline{\textbf{Step 4:}} The linear value corresponding to the rank determined in Step 3 is calculated using.
\begin{equation}
    i = \lfloor \mbox{Rank} \rfloor, ~~~~~d = \mbox{Rank} - i,
\end{equation}
where $i$ is the integer part and $d$ is the decimal part. As such, the new percentage can be computed using
\begin{equation}
    p = x_i + d (x_{i+1} - x_i).    
\end{equation}
By applying these procedures, a newly optimised tolerance value can be obtained \cite{Bornmann}. \vspace{1mm}   

\noindent \underline{\textbf{Step 5:}} It is crucial to ensure that all true negatives (actual defects) continue to be flagged even when the newly optimised tolerance value is used.

\begin{figure}[t]
    \centering
    \includegraphics[width=\columnwidth]{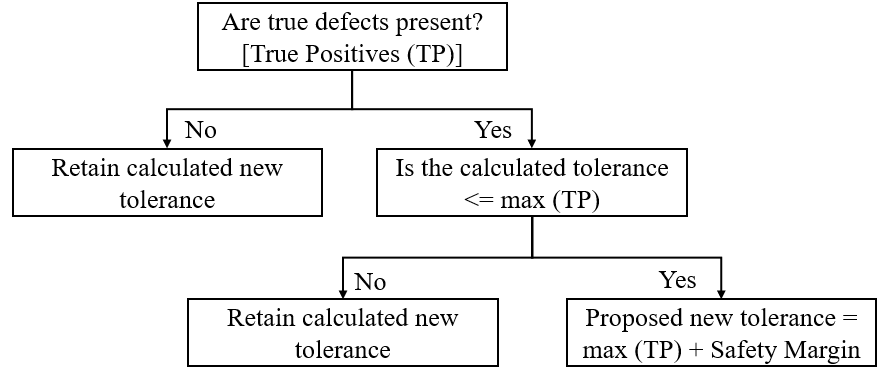}
    \caption{The process to ensure all true defects are flagged as seen in Step 5.}
    \label{fig:fig2}
\end{figure}

This is achieved using the illustration shown in Fig. \ref{fig:fig2}. If the dataset contains no actual defects, the computed tolerance value is retained. However, if actual defects are present, the maximum actual defect value is compared with the tolerance. This is because once the defect value closest to the tolerance (i.e., the maximum actual defect value) is flagged, all other defect values will also be flagged. If the system fails to detect an actual defect, the existing tolerance value is discarded and a new tolerance value is defined by adding a safety margin to the maximum defect value. This procedure ensures all genuine defects are still flagged to eliminate risks of increasing true defect escapes, while also reducing the number of false calls. %\vspace{1mm}

\vspace{-3mm}
\noindent \underline{\textbf{Step 6:}} This is the validation stage. It is essential to confirm that all true defects are flagged. To achieve this, validation is conducted using previously unseen data. The validation set, reserved earlier, is utilised to verify whether the model flags all true defects under the newly optimised tolerance.

We employ the concept of percentile rank instead of averaging, as averages can be heavily influenced by outliers. Consider an example of five inspection values: 2, 28, 35, 32, and 25, with the existing tolerance being 40. Averaging these values yields a new tolerance of 24.4, which is a substantial and undesirable reduction from the existing value. In contrast, calculating the percentile rank for these values results in a tolerance of 33.2, which is more reliable. Employing percentiles also offers flexibility in adjusting the extent to which the tolerance is modified. One can simply alter the percentile threshold and simulate potential outcomes before implementing any changes on the actual AOI system. This demonstrates the advantages of percentile rank over average in determining tolerance values.

\section{Results and Discussion}\label{sec:results}
Proof of concept was obtained through preliminary research, using data from the solder inspections of 917 different parts. Collectively, these parts exhibited 54,019 false flags and 544 true defects. Solder inspection is conducted in the AOI machine by analysing the reflection characteristics of light projected onto the solder joints. Considering a percentile rank of 80, new tolerances for each part were generated, resulting in a reduction of false flags to 44,209 while retaining the 544 true defects. This corresponds to an 18\% decrease in false flags, a notable yet reliable improvement. Meanwhile, using a 75\% percentile rank yields 42,142 false flags (a 22\% decrease) while maintaining the same count of true defects. In scenarios requiring higher reliability, a higher percentile rank is recommended. Beneficially, our simulation capabilities allow us to assess the effectiveness of new percentile values before implementing them in the AOI machines.

As shown in the sample data in Table \ref{tab:tab2}, all true defects have been successfully flagged even with the implementation of the new tolerance. Proprietary details, including model names and part numbers, have been removed to ensure confidentiality.
\begin{table*}[t]
    \centering
    \caption{Sample set of data.} \vspace{-2mm}
    \label{tab:tab2}
    \begin{tabular}{@{}cccccccc@{}}
    \toprule
    \textbf{Model} &
      \textbf{\begin{tabular}[c]{@{}c@{}}Part \\ Number\end{tabular}} &
      \textbf{\begin{tabular}[c]{@{}c@{}}Current \\ Tolerance\end{tabular}} &
      \textbf{\begin{tabular}[c]{@{}c@{}}False Call \\ Count\end{tabular}} &
      \textbf{\begin{tabular}[c]{@{}c@{}}True Defect \\ Count\end{tabular}} &
      \textbf{\begin{tabular}[c]{@{}c@{}}Optimised \\ Tolerance\end{tabular}} &
      \textbf{\begin{tabular}[c]{@{}c@{}}New False \\ Call Count\end{tabular}} &
      \textbf{\begin{tabular}[c]{@{}c@{}}New True \\ Defect Count\end{tabular}} \\ \midrule
    A & P1 & 42.62 & 2264 & 5  & 40.50 & 1885 & 5  \\
    B & P2 & 45.00 & 1756 & 0  & 42.30& 1392 & 0  \\
    C & P3 & 40.00 & 1231 & 2  & 35.00 & 984  & 2  \\
    D & P4 & 30.00 & 1100 & 3  & 27.60 & 879  & 3  \\
    E & P5 & 45.00 & 25   & 27 & 44.20 & 19   & 27 \\
    F & P6 & 45.00 & 605  & 5  & 43.30 & 478  & 5  \\
    G & P7 & 25.00 & 13   & 5  & 23.68 & 10   & 5  \\ \bottomrule
    \end{tabular}
\end{table*}

\begin{figure}[t]
    \centering
    \includegraphics[width=\columnwidth]{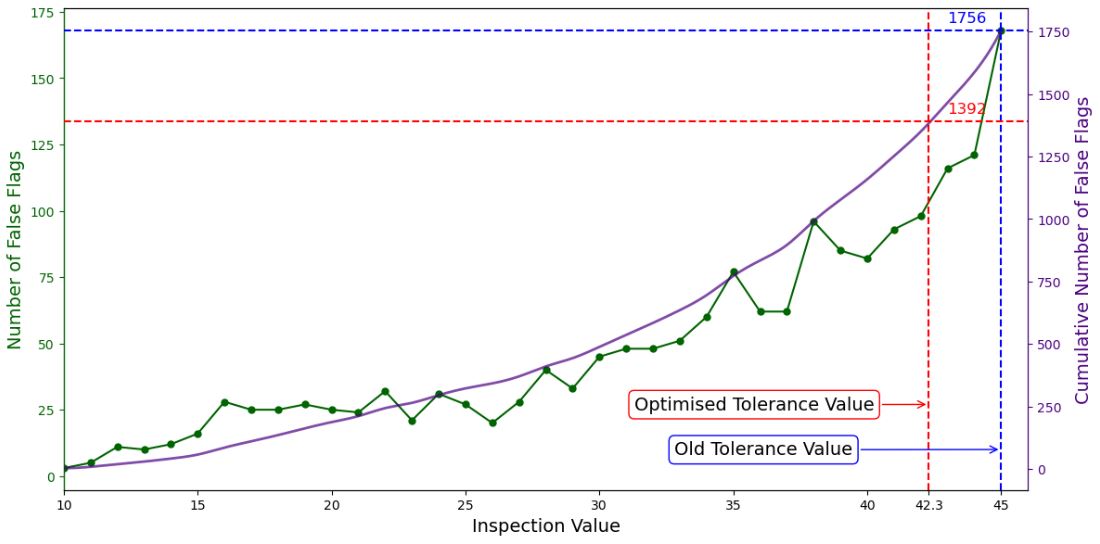}

    \caption{False flags before and after optimisation for part P2 from Table \ref{tab:tab2}.}
    \label{fig:fig3}
\end{figure}

Fig. \ref{fig:fig3} depicts the distribution of false flag inspection values together with both original and optimised tolerances for part P2 (see Table \ref{tab:tab2}). With the initial tolerance of 45, this component generated 1756 false flags, and adopting the new tolerance of 42.3 (corresponding to the 80\% percentile) reduces that figure to 1392 --- a decrease of 364 false flags, or 20.7\%.

\begin{table*}[t]
    \centering
    \caption{Output of the validation process.} \vspace{-2mm}
    \label{tab:tab3}
    \begin{tabular}{@{}ccccccc@{}}
        \toprule
        \textbf{Model} &
          \textbf{Part Number} &
          \textbf{\begin{tabular}[c]{@{}c@{}}Training True \\ Defect Count\end{tabular}} &
          \textbf{\begin{tabular}[c]{@{}c@{}}Validation True \\ Defect Count\end{tabular}} &
          \textbf{\begin{tabular}[c]{@{}c@{}}True Defect \\ Flagged\end{tabular}} &
          \textbf{\begin{tabular}[c]{@{}c@{}}True Defects \\ Escaped\end{tabular}} &
          \textbf{Validation Status} \\ \midrule
        A & V1 & 28 & 13 & 13 & 0 & Pass \\
        B & V2 & 81 & 35 & 35 & 0 & Pass \\
        C & V3 & 18 & 9  & 9  & 0 & Pass \\
        D & V4 & 22 & 10 & 10 & 0 & Pass \\
        E & V5 & 23 & 11 & 11 & 0 & Pass \\ \bottomrule
    \end{tabular}
\end{table*}
The framework was subsequently validated on previously unseen data, and the results are presented in Table \ref{tab:tab3}. Hence, the recall can be computed using 
\begin{equation} 
    \mbox{Recall} = \frac{TP}{TP + FN}, \label{eq:recall}
\end{equation} 
where $TP$ and $FN$ represent the true positives and false negatives, respectively. 

Using (\ref{eq:recall}), a recall of 100\% was achieved, showing that all true defects have been captured. This outcome aligns with our primary objective and confirms the reliability of the system. The findings indicate promising results from the proposed tolerance optimisation technique, where an 18\% reduction in false calls was achieved at the 80\% percentile.
% , while maintaining a 100\% recall rate to ensure all true defects continue to be flagged. 
Further model development is underway to incorporate real-time tolerance data captured directly from AOI machines, facilitating continuous improvement. The accuracy of the model is expected to increase as the dataset expands. Ultimately, the model is anticipated to reach a performance level where false calls are effectively eliminated, and all flagged instances represent genuine defects, thereby removing the need for manual inspection. This work underpins the creation of a digital shadow, or twin, to provide insights into the performance of the physical system under varying operating conditions and facilitating predictive maintenance. See, for example \cite{wucherer2023, wuchererCCTA}.

In addition to reducing the number of false calls, significant emphasis is placed on ensuring that the optimisation system remains reliable, particularly for critical applications. The system is designed for ease of accessibility and seamless integration into existing industrial processes, minimising implementation risks. Furthermore, it can be readily applied to new products introduced into the production line, offering flexibility and adaptability.

% \section*{Acknowledgment}
% This project is funded by Innovate UK Knowledge Transfer Network (KTN) and Invest Northern Ireland (Invest NI) [Project Number: 10078007].

\section{Conclusion}\label{sec:conclusion}
Overall, the proposed tolerance optimisation framework demonstrates substantial potential to enhance AOI processes within semiconductor manufacturing, especially for SMT. By systematically reducing false calls and ensuring all genuine defects are reliably detected, this approach contributes significantly towards improved efficiency, reduced operational costs, and higher quality standards for products. This framework is readily integrable with the AOIs currently used in the industry. Future work will focus on further refining the system to incorporate broader real-time data integration and exploring its scalability to other critical manufacturing sectors.

\bibliographystyle{IEEEtran}
\bibliography{IEEEabrv,ref}

\begin{thebibliography}{1}
\providecommand{\url}[1]{#1}
\csname url@samestyle\endcsname
\providecommand{\newblock}{\relax}
\providecommand{\bibinfo}[2]{#2}
\providecommand{\BIBentrySTDinterwordspacing}{\spaceskip=0pt\relax}
\providecommand{\BIBentryALTinterwordstretchfactor}{4}
\providecommand{\BIBentryALTinterwordspacing}{\spaceskip=\fontdimen2\font plus
\BIBentryALTinterwordstretchfactor\fontdimen3\font minus \fontdimen4\font\relax}
\providecommand{\BIBforeignlanguage}[2]{{%
\expandafter\ifx\csname l@#1\endcsname\relax
\typeout{** WARNING: IEEEtran.bst: No hyphenation pattern has been}%
\typeout{** loaded for the language `#1'. Using the pattern for}%
\typeout{** the default language instead.}%
\else
\language=\csname l@#1\endcsname
\fi
#2}}
\providecommand{\BIBdecl}{\relax}
\BIBdecl

\bibitem{yimin2023industrial}
W.~Yimin and X.~Yu, ``Industrial chip positioning method in surface mount technology,'' in \emph{IECON 2023-49th Annual Conference of the IEEE Industrial Electronics Society}.\hskip 1em plus 0.5em minus 0.4em\relax IEEE, 2023, pp. 1--4.

\bibitem{Kim}
D.~Kim, J.~Koo, H.~Kim, S.~Kang, S.~H. Lee, and J.~T. Kang, ``{Rapid fault cause identification in surface mount technology processes based on factory-wide data analysis},'' \emph{International Journal of Distributed Sensor Networks}, vol.~15, no.~2, 2019.

\bibitem{Lim}
K.~Y.~H. Lim, P.~Zheng, and C.-H. Chen, ``{A state-of-the-art survey of Digital Twin: techniques, engineering product lifecycle management and business innovation perspectives},'' \emph{Journal of Intelligent Manufacturing}, vol.~31, no.~6, pp. 1313--1337, 2020.

\bibitem{Manjurul}
M.~M.~M. Islam, J.~I. Emon, K.~Y. Ng, A.~Asadpour, M.~M. R.~A. Aziz, M.~L. Baptista, and J.-M. Kim, \emph{Artificial Intelligence in Smart Manufacturing: Emerging Opportunities and Prospects}.\hskip 1em plus 0.5em minus 0.4em\relax Springer Nature Switzerland, 2025, pp. 9--36.

\bibitem{Res}
V.~Reshadat and R.~A. Kapteijns, ``{Improving the Performance of Automated Optical Inspection (AOI) Using Machine Learning Classifiers},'' in \emph{2021 International Conference on Data and Software Engineering (ICoDSE)}, 2021, pp. 1--5.

\bibitem{9935366}
I.~Jamal, M.~A. Nasir, C.~Ani Adi~Izhar, M.~Maruzuki, and K.~Ishak, ``Fine-tuning strategy for re-classification of false call in automated optical inspection post reflow,'' in \emph{2022 2nd International Conference on Emerging Smart Technologies and Applications (eSmarTA)}, 2022, pp. 1--5.

\bibitem{Bornmann}
L.~Bornmann, L.~Leydesdorff, and J.~Wang, ``{Which percentile-based approach should be preferred for calculating normalized citation impact values? An empirical comparison of five approaches including a newly developed citation-rank approach (P100)},'' \emph{Journal of Informetrics}, vol.~7, no.~4, pp. 933--944, 2013.

\bibitem{wucherer2023}
\BIBentryALTinterwordspacing
S.~Wucherer, R.~McMurray, K.~Y. Ng, and F.~Kerber, ``{Learning to Predict Grip Quality from Simulation: Establishing a Digital Twin to Generate Simulated Data for a Grip Stability Metric},'' 2023. [Online]. Available: \url{https://arxiv.org/abs/2302.03504}
\BIBentrySTDinterwordspacing

\bibitem{wuchererCCTA}
------, ``Predicting maximum permitted process forces for object grasping and manipulation using a deep learning regression model,'' in \emph{2024 IEEE Conference on Control Technology and Applications (CCTA)}, 2024, pp. 669--674.

\end{thebibliography}

\end{document}